\documentclass[aps,prd,superscriptaddress,twoside,twocolumn,nofootinbib,%
showpacs,floatfix]{revtex4-1}

\usepackage{amsmath,amssymb}
\usepackage{graphicx,bm}
\usepackage{slashed}

\usepackage{ulem} %% for strike-through
\usepackage[usenames]{color}

\renewcommand\sout{\bgroup \color{red} \ULdepth=-.5ex \ULset}

\begin{document}

\title{Hidden Local Symmetry and Infinite Tower of Vector Mesons
for Baryons}
\vskip 1cm

\author{Yong-Liang Ma}
\email{ylma@hken.phys.nagoya-u.ac.jp}
\affiliation{Department of Physics,  Nagoya University,  Nagoya,
464-8602, Japan}

\author{Yongseok Oh}
\email{yohphy@knu.ac.kr}
\affiliation{Department of Physics,
Kyungpook National University, Daegu 702-701, Korea}

\author{Ghil-Seok Yang}
\email{ghsyang@gmail.com}
\affiliation{Department of Physics,
Kyungpook National University, Daegu 702-701, Korea}

\author{Masayasu~Harada}
\email{harada@hken.phys.nagoya-u.ac.jp}
\affiliation{Department of Physics,  Nagoya University,  Nagoya,
464-8602, Japan}

\author{Hyun Kyu Lee}
\email{hyunkyu@hanyang.ac.kr}
\affiliation{
Department of Physics, Hanyang University, Seoul 133-791, Korea}

\author{Byung-Yoon Park}
\email{bypark@cnu.ac.kr}
\affiliation{Department of Physics, Chungnam National University, Daejeon 305-764, Korea}

\author{Mannque Rho}
\email{mannque.rho@cea.fr}
\affiliation{Institut de Physique Th\'eorique,
CEA Saclay, 91191 Gif-sur-Yvette c\'edex, France \& \\
Department of Physics, Hanyang University, Seoul 133-791, Korea}

\date{\today}

\begin{abstract}
In an effort to access dense baryonic matter relevant for compact stars in a unified framework
that handles both single baryon and multibaryon systems on the same footing, we first address
a holographic dual action for a single baryon focusing on the role of the infinite tower of vector
mesons deconstructed from five dimensions. To leading order in 't Hooft coupling
$\lambda=N_c g_{\rm YM}^2$, one has the Bogomol'nyi-Prasad-Sommerfield (BPS) Skyrmion
that results when the warping of the bulk background and the Chern-Simons term in the
Sakai-Sugimoto D4/D8-${\overline {\rm D8}}$ model are ignored.
The infinite tower was found by Sutcliffe to induce flow to a conformal theory,
i.e., the BPS.  We compare this structure to that of the SS model consisting of a 5D Yang-Mills
action in warped space and the Chern-Simons term in which higher vector mesons are integrated
out while preserving hidden local symmetry and valid to $O(\lambda^0)$ and $O(p^4)$
in the chiral counting.
We point out the surprisingly important role of the $\omega$ meson that figures in the Chern-Simons
term that encodes chiral anomaly in the baryon structure and that may be closely tied to short-range
repulsion in nuclear interactions.
\end{abstract}

\pacs{
12.39.Dc,	% Skyrmions
12.39.Fe,	% Chiral Lagrangians
14.20.-c	% Baryons
}

\maketitle

%%%%%%%%%%%%%%%%%%%%%%%%%%%%%%%%%%%%%%%%%%%%%%%%%%%%%

\section{Introduction}
There is a growing evidence that the infinite tower of vector mesons play an important role for
the baryon structure and consequently for dense baryonic matter.
From the theoretical point of view, there is a natural rationale for their role,
both bottom-up and top-down.

At very low energy, Quantum Chromodynamics (QCD) is effectively modeled by
nonlinear sigma model encapsulating current algebra and as the energy scale increases,
there emerge massive vector excitations. An elegant way of capturing the physics of
vector mesons is to exploit that there are redundancies in the chiral field representing
the pseudo-Goldstone bosons, pions, and introduce gauge symmetry associated
with the redundancies. The nonlinear sigma model is gauge-equivalent to hidden local symmetry
(HLS)~\cite{BKY88,HY03a}, so the vector mesons so generated can be identified with
the hidden local gauge fields.
In fact, there are an infinite number of redundancies as the energy goes up and hence an
infinite number of gauge fields. The infinite number of hidden gauge vector fields together
with the pion field in 4D can be dimensionally de-constructed to 5D Yang-Mills (YM) action
in curved space~\cite{SS03}.
Here the 5th dimension represents energy scale. This is referred to as ``bottom-up" approach.

A similar 5D YM structure arises in the gravity sector (that is referred to as ``bulk" sector) of
gravity/gauge (holographic) duality that comes from string theory.
Among a variety of models given in the bulk sector, referred to as holographic models,
the one which has the properties closest to QCD is the model constructed by
Sakai and Sugimoto (SS) using D4/D8-$\overline{\rm D8}$ branes~\cite{SS04a-SS05}.
When Kaluza-Klein(KK)-decomposed to 4D, this model gives an infinite tower of vector mesons
plus the pions which map to those of the dimensionally de-constructed gauge theory given on
the boundary.
This dual (bulk-sector) model is justified in the large $N_c$ and large 't Hooft
$\lambda = N_c g^2_{\rm YM}$
limit and the chiral limit where the quark masses are taken to be zero.
In these limits, there are only two parameters in the model and they are fixed from meson dynamics.
We call this ``top-down" approach.

This paper is the first in the series of studies made to  arrive at a description of  dense
baryonic matter in a unified scheme in which both single baryon and mutl-baryons are
treated on the same footing.
In this paper which will focus on the single baryon properties, we will simply adopt the SS model
in developing the dynamics of baryons which will in subsequent publications be applied to
many-baryon systems, including dense baryonic matter relevant to compact stars.
Given the three limits adopted, large $N_c$, large $\lambda$ and chiral limit,  which do not
always apply in nuclear dynamics, the model cannot be expected to work well for
\textit{all} baryonic properties and processes, but the merit of this model is that one can make
a precise set of parameter-free calculations that have not been done in the past in the field.
Such a feat is made feasible because there are no unknown parameters once they are fixed
in the meson sector.
For the single baryon, regardless of how well it fares with Nature, this could be taken as a
land-mark calculation in that it is the first \textit{complete and parameter-free} soliton calculation
with a chiral Lagrangian with vector mesons written up to $O(p^4)$ including all of
the homogeneous Wess-Zumino terms.

Up to date, there is no workable model-independent theoretical tool available to treat
simultaneously the structure of elementary baryon and many-baryon systems
(such as nuclei and nuclear matter).
Lattice QCD cannot access dense matter because of the sign problem which remains unresolved.
One possible approach that unifies both elementary baryons and multi-baryon systems was
proposed in Ref.~\cite{LPMRV03} where starting with a chiral Lagrangian, the single baryon is
generated as a Skyrmion and multi-Skyrmions are put on crystal lattice to simulate many-baryon
systems and dense matter.
In this series of work, we apply the same strategy with the Lagrangian having the infinite-tower 
of vector mesons that arises either from string theory or dimensionally deconstructed theory to 
both nucleon structure and dense matter. 
The former is treated here and the latter will be given in a forthcoming publication.%
\footnote{There have been works that incorporate vector mesons and other degrees of freedom 
in calculating properties of the single Skyrmon~\cite{meissner} and dense 
baryonic matter~\cite{park-vento}. There have also been detailed structure calculations of 
few-Skyrmion systems using the Skyrme model~\cite{manton-sutcliffe}. 
As will be stressed throughout the paper, what distinguishes the work(s) described in the present paper 
from the previous works is that once the pion decay constant $f_\pi$ and the $\rho$-meson mass 
$m_\rho^{}$ are fixed from the meson sector, this work is the first truly parameter-free treatment of 
single Skyrmion as well as multi-Skyrmions with a hidden local symmetry Lagrangian valid to 
chiral $O(p^4)$ and in the large $N_c$ and 't Hooft constant limit.}

To start with, we motivate our development with the observation made by Sutcliffe~\cite{sutcliffe11}
on the structure of Skyrmions when the warping of the holographic direction and the
Chern-Simons term are turned off, which amounts to taking the large $\lambda$ limit, that is,
keeping only the $O(\lambda)$ terms.
The resulting Skyrmion is a BPS, that is, a conformally invariant object, to which, it is found,
the theory flows as more and more of the infinite tower of vector mesons in 4D enter.
Of course the BPS Skyrmion by itself is trivial in the sense that there is no interaction in it:
It cannot capture the physics of Nature which has interactions.
Therefore it is the \textit{deviation from BPS}, namely, the warping of the background \textit{and}
the Chern-Simons term both of which enter at $O(\lambda^0)$  that encodes nontrivial physics.
We show how this feature arises by means of truncating the SS model with the lowest-lying
vector mesons $\mathcal{V}_1\equiv (\rho,\omega)$.
We will see that the $U(1)$ degree of freedom residing in the Chern-Simons term, namely the
$\omega$ meson, that prevents the soliton from shrinking~\cite{HRYY07-HRYY07a-HRYY07b,HSSY07}, not only blocks
the flow to conformal fixed point but also plays a very important role in the Skyrmion structure
of baryons and consequently in nuclear many-body interactions, i.e., dense matter.
It will also be seen that there is a crucial need for a low-mass scalar -- which is famously missing --
in the top-down holographic model in a way analogous to what happens in the mean-field model
of nuclear matter.
In nuclear matter, the small binding energy $\sim 16$ MeV arises from a nearly exact cancelation
between the $\omega$ repulsion and the attraction due to a scalar of mass comparable to that of
the $\omega$. We conjecture that a similar phenomenon is taking place in the dynamics for both
single Skyrmion and multi-Skyrmion systems.

\section{The Holographic Model}

We start with the holographic action derived by Sakai and Sugimoto in the large $N_c$ and
$\lambda$ limit.
For our purpose, it is not necessary to enter into the details of how the action is derived from
the gravity-gauge duality in string theory.
It suffices for our purpose to state simply that it gives the generic structure of 5D YM action
with no free parameters that is holographically dual to what corresponds to QCD in the large
$N_c$ and $\lambda$ limit (and the chiral limit).
As such it can be reliable for certain quantities where $1/N_c$ and/or $1/\lambda$ corrections
are unimportant but not for certain others.
The holographic dual action of the SS model~\cite{SS04a-SS05} can be written after a suitable redefinition
in the form~\cite{HRYY07-HRYY07a-HRYY07b,KMS12}
\begin{equation}
S = S_{\rm DBI} + S_{\rm CS}
\end{equation}
where
\begin{equation}
S_{\rm DBI} \approx S_{\rm YM} = -\kappa \int d^4x dz \frac{1}{2e(z)^2}
\, \mbox{tr}\, \mathcal{F}_{mn}^2
\label{dbiaction}
\end{equation}
with $\kappa=\frac{\lambda N_c}{216\pi^3}$, $e(z)$ is the effective YM coupling that depends
on the holographic direction $z$ and is proportional to the KK mass as $M_{KK}^{-1/2}$ and
$S_{\rm CS}$ is the Chern-Simons (CS) action that comes from the coupling of the D8-branes
to the bulk Ramond-Ramond field.
We use the index $m=(\mu,z)$ with $\mu=0,1,2,3$.
The gravity enters in the $z$ dependence of the YM coupling, giving rise to the warping of the space.
$\mathcal{A} = \mathcal{A}_\mu dx^\mu + \mathcal{A}_z dz$ is the five-dimensional
U($N_f$) gauge field and $\mathcal{F} = d \mathcal{A} + i \mathcal{A} \mathcal{A}$ is its field strength.
We are interested in $N_f=2$, so the gauge field is
\begin{equation}
\mathcal{A}=A_{\rm SU(2)}+\frac 12 \tilde{A}_{\rm U(1)}.
\end{equation}
For this the YM term is
\begin{equation}
S_{\rm YM}=-\kappa\int d^4x dz\frac{1}{2e^2(z)} \left( \,\mbox{tr}\, F_{mn}^2 +
\frac 12 \tilde{F}_{mn}^2 \right)
\label{ym}
\end{equation}
and the CS term
\begin{equation}
S_{\rm CS}=\frac{N_c}{16\pi^2} \int\tilde{A} \wedge \mbox{tr} F^2 +
\frac{N_c}{96\pi^2}\int {\tilde A}\wedge {\tilde{F}}^2.
\label{CS}
\end{equation}
In Eqs.~(\ref{ym}) and (\ref{CS}) $F_{mn}$ is the field strength for the SU(2) gauge field
and $\tilde{F}_{mn}$ stands for the field strength of the U(1) gauge field.
Back-reactions are ignored in these expressions.

To the leading order in $\lambda$, that is, to $O(\lambda)$, $e(z)$ is a constant,
so the 5D YM action can be taken to be in flat space.
In fact one can ignore the $O (\lambda^0)$ contribution in computing static energy,
so up to $O (\lambda^0)$, the static baryon is given by the instanton solution that is
self-dual~\cite{KMS12}.

\section{BPS Skyrmion}

The role of the infinite tower of vector mesons in the baryon structure can be studied in the
approximation that the space is flat in Eq.~(\ref{dbiaction}) and the CS term is ignored.
This corresponds to taking the leading $O(\lambda)$ in the SS action.
This looks like a drastic approximation as we will see later.
In particular, the ignoring of the CS term, although subleading in $\lambda$,
is found to be suspect for nucleon structure.
However it can give us a good idea of how the infinite tower encoded in the 5D YM action
figures in the nucleon structure as well as in dense medium.
The Skyrmion of this action, called BPS Skyrmion, was studied by Sutcliffe~\cite{sutcliffe11,sutcliffe10}.
We first review this model because it illustrates clearly the kind of physics we would like to explore.
We will uncover the role of the lowest vector mesons $\rho$ and $\omega$ and the effect of the
higher members in the structure of both elementary and multi-body systems.

As with Sutcliffe, we consider the 5D Euclidean YM action%
\footnote{This is in unit of an arbitrary mass dimension, so the energy discussed below in this section is in that unit.
In the sections that follow with the SS model, the coefficient will be specified.}
\begin{equation}
S=-\frac 12\int \mbox{tr} F_{mn}^2 d^4x\, dz ,
\label{BPSaction}
\end{equation}
where
\begin{equation}
F_{mn} = \partial_m A_n-\partial_n A_m +[A_m,A_n]
\end{equation}
with $A_m = T^a A^a_m$ normalized $\mbox{tr}(T^a T^b) = \frac 12 \delta_{ab}$.
The gauge field transforms
\begin{equation}
A_m \rightarrow g (A_m+\partial_m) g^{-1}.
\end{equation}
The static energy coming from the action (\ref{BPSaction}), known as BPS action,
has a well-known bound, the Bogomolnyi bound,%
\footnote{This bound differs from Sutcliffe's expression by a factor of 4
because Sutcliffe's $B$ seems to be 4 times our definition in Eq.~(\ref{B}).}
\begin{equation}
E \geq 8\pi^2 B
\label{bound}
\end{equation}
with
\begin{equation}
B = \frac{1}{16\pi^2} \int \mbox{tr} (F_{MN} {}^*F_{MN}) d^3 x \, dz
\label{B}
\end{equation}
where $M=1,2,3,z$ and ${}^*F_{MN}=\frac 12\epsilon_{MNAB}^{} F_{AB}^{}$ is the dual field strength.
Now the bound is satisfied if $F_{MN}$ is self-dual, i.e.,
\begin{equation}
F_{MN} = {}^*F_{MN}.
\label{selfdual}
\end{equation}
This means that the energy of the system cannot be lower than the bound.

In order to see how the 4D meson fields that are measured in the laboratories enter into
the theory, one needs to do the mode expansion,
\begin{eqnarray}
A_\mu (x^\mu,z) &=& \sum_{n\geq 1} V^n_\mu(x^\mu) \psi_n(z),
\nonumber\\
A_z(x^\mu,z) &=& \sum_{n\geq 0} \varphi^n(x^\mu) \phi_n(z).
\label{modeexpansion0}
\end{eqnarray}
We work with the gauge $A_z=0$ which can be obtained by taking
\begin{equation}
g(\mathbf{x},z) = \mathcal{P} \exp  \int_{0}^z A_z(\mathbf{x},z^\prime) \, dz^\prime .
\label{gaugechoice}
\end{equation}
In the new gauge with the gauge-transformed field $A^g_z=0$, with the requirement that
$A_m\rightarrow 0$ for $|z| \rightarrow \infty$, we have (in the absence of external fields)
\begin{equation}
A_i^g\rightarrow -\xi_{R,L}^{} \partial_i \xi_{R,L}^{-1} \equiv \alpha_{i}^{R,L},\, \ z\rightarrow \pm \infty
\end{equation}
where
\begin{equation}
\xi_{R,L}^{}(\mathbf{r}) \equiv g(\mathbf{x}, \pm \infty).
\label{anz}
\end{equation}
This shows that the chiral field $U \equiv \xi_L^\dag \cdot \xi_R = e^{if(r) \bm{\tau} \cdot \bm{\pi} }$
appears at the boundary -- with the external fields turned off -- and is given by the holonomy as in
the Atiyah-Manton ansatz~\cite{AM89}.%
\footnote{Note that in the case of Atiyah and Manton, the holonomy is in the time direction
while here it is in the fifth ($z$) direction.}

Then gauge-transformed mode expansion (\ref{modeexpansion0}) takes the form
\begin{eqnarray}
A_\mu^g(x^\mu,z) & = & \alpha_\mu^R (x^\mu)\phi^R(z) + \alpha_\mu^L (x^\mu)\phi^L(z)
\nonumber\\ & & \mbox{}
+ \sum_{n\geq 1} \left[ A_\mu^n(x^\mu)\psi_{2n}(z) - V_\mu^n(x^\mu)\psi_{2n-1}(z) \right].
\nonumber\\
\label{modeexpansion}
\end{eqnarray}
Here $V_\mu^n(x^\mu)$ and $A_\mu^{n}(x^\mu)$ are the vector and axial-vector meson fields,
respectively, and $\psi_n$ is a function that satisfies the equation
\begin{equation}
-\partial_z^2 \psi_n (z) = \lambda_n \psi_n . \label{eq:eigenflat}
\end{equation}
Note that this eigenvalue equation by itself has plane wave solutions and continuous spectra. 
However, in the present case, Eq.~(\ref{eq:eigenflat}) is subject to the requirement that the solution
be a complete orthonormal basis for square integrable functions on the real line with unit weight 
functions, which is necessary to obtain canonical kinetic terms for the vector mesons~\cite{sutcliffe10}. 
This requirement leads to a Hermite function
\begin{equation}
\psi_n(z)=\frac{(-1)^n}{\sqrt{n!\,2^n\sqrt{\pi}}}e^{\frac{1}{2}z^2}\frac{d^n}{dz^n}e^{-z^2}
\end{equation}
normalized as
\begin{equation}
\int_{-\infty}^{\infty} \psi_m(z)\psi_n(z)\, dz=\delta_{mn}.
\end{equation}
This allows to do the $z$ integration, so the problem reduces to 4D.
With the Hermite function, we have
\begin{eqnarray}
\phi^{R,L}(z) & = &
\frac{1}{\sqrt{2}\pi^{1/4}} \int_{\mp\infty}^{\pm z} \psi_0^{} (\xi)\, d\xi\nonumber\\
&=&\frac{1}{2} \pm \frac{1}{2} \, \mbox{erf}(z/\sqrt{2}),
\label{psiplus}
\end{eqnarray}
where $\mbox{erf}(z)$ is the usual error function $\mbox{erf}(z)=\frac{2}{\sqrt{\pi}}\int_0^z e^{-\xi^2}\, d\xi$.
The normalization of $\phi^{R,L}(z)$ is chosen so that
$\phi^{R,L}(\mp\infty)=0$ and $\phi^{R,L}(\pm \infty)=1$.

What we are interested in is how the tower of vector mesons contributes to the static energy of the action given in Eq.~(\ref{BPSaction}).
Briefly the important observation made by Sutcliffe is this.
The more vector mesons are included, the closer the static energy goes down and approaches the BPS bound.
In other words, the higher tower of vector mesons drive the theory to a conformal theory.

In order to explore the role of the tower, first consider eliminating all the vector mesons and
leave only the pions as the explicit degrees of freedom.
Written in terms of the tower of hidden local gauge fields as is explained in Refs.~\cite{SS04a-SS05},
this can be done by ``integrating out" the (hidden local) gauge fields.
Then one winds up with the energy of the Skyrme model with the current algebra term and
an ``effective" or renormalized Skyrme quartic term~\cite{Skyrme62}
\begin{eqnarray}
E^{(0)} &=& \int \left( \frac{C_1^{}}{2} \mbox{tr} (\partial_\mu U^\dagger \partial^\mu U)
\right. \nonumber \\ && \mbox{} \qquad \left.
+ \frac{C_2^{}}{16} \mbox{tr} \left[U^\dagger\partial_\mu U,U^\dagger\partial_\nu U \right]^2
\right) \, d^3x,
\label{e0}
\end{eqnarray}
where $C_i$'s are constants given by the integral over the Hermite polynomials
and $U$ is given by the ``holonomy" in Eq.~(\ref{anz}),
\begin{equation}
U(\mathbf{x}) = \mathcal{P} \exp\,\int_{-\infty}^\infty A_z(\mathbf{x},z^\prime)dz^\prime.
\label{U-AM}
\end{equation}
One can calculate the energy of the soliton by using an instanton ansatz as in Atiyah-Manton~\cite{AM89}
or in the exact numerical way~\cite{JR83,ANW83}. They give very close results
\begin{equation}
E^{(0)}= 1.235\, (8\pi^2 B).
\label{1skyrmion}
\end{equation}
This is the usual 1.24 times the bound, here the Bogomol'nyi bound (\ref{bound})
which corresponds in the case of the Skyrme Lagrangian to the Faddeev bound $12\pi^2 B$.

\subsection{The infinite tower and conformal symmetry}

Now what happens when the vector mesons are included?
There are no free parameters so this question can be answered precisely.
The result is quite striking.
As shown by Sutcliffe, the lowest lying vector meson $\rho$ brings the energy
from  Eq.~(\ref{1skyrmion}) down to
\begin{equation}
E^{(1)}=1.071(8\pi^2 B)
\end{equation}
and the next-lying axial-vector meson $a_1^{}$ brings this further down to
\begin{equation}
E^{(2)}=1.048(8\pi^2 B).
\end{equation}
Since the full tower will bring this to the bound $E^{(\infty)}=8\pi^2B$,
it follows that the high-lying vector mesons make the theory flow to a conformal theory.
That the lowest-lying vector meson does nearly all the work in flowing to the conformality is
reminiscent of the near complete saturation of the charge sum rule%
\footnote{In fact it overshoots the charge.}
of the pion~\cite{SS04a-SS05} and nucleon~\cite{HRYY07-HRYY07a-HRYY07b,HSS08} form factors.

A very analogous tendency is seen when the BPS model is applied to finite nuclei:
the vector mesons mediate the flow to conformality and furthermore, reduce the over-binding of nuclei
in the Skyrme model~\cite{sutcliffe11}.

\subsection{\boldmath The $\omega$ meson and the Chern-Simons term}

As stated, the BPS Skyrmion considered above is strictly justified in the large $\lambda$ limit
(in addition to the large $N_c$ and chiral limits).
To next order in $\lambda$, the metric is curved in the holographic direction.
To that order,  the Chern-Simons term enters bringing in a U(1) degree of freedom, i.e.,
the $\omega$ meson and its tower.
In fact the entire tower gives rise to the universal $1/r^2$ repulsion in the holographic model~\cite{AHI12}.
We know from nuclear physics that the $\omega$ meson brings in repulsion, without which nuclei will collapse.
In  the Skyrmion description, what it does is to make the soliton mass appreciably increased compared
with the one without it~\cite{PRV03}.
In nuclei, the binding requires the presence of a scalar, say, $\phi$
(often denoted as $\sigma$ -- which is not the fourth component of the chiral four-vector in sigma models).
It is the near cancellation of the $\omega$ repulsion and the scalar attraction that gives the small binding energy
of nuclear matter $\sim 16$~MeV.

It is clear from the above consideration that both the warping of the background deviating from the BPS structure
and the Chern-Simons term needs to be confronted.
This means that we have to address the infinite tower structure in the presence of warping and the Chern-Simons term
including all the terms to $O(\lambda^0)$ and chiral order $O(p^4)$.
This problem has been worked out fully in a highly involved calculation with no free parameters in Ref.~\cite{MOYH12}.
Here we use their results to show certain intricately contrary roles played by the iso-vector and iso-scalar vector mesons
in the baryon structure and make conjectures on their potential influence in dense matter.

\section{Integrating-out of the tower of vector mesons}

We return to the Sakai-Sugimoto model in its original form%
\footnote{Here, we only keep the leading terms in the $1/\lambda$ expansion. 
The inclusion of other terms in the $1/\lambda$ expansion will introduce more terms in the action such as the $F^3$ terms. 
But the contribution from these non-leading $1/\lambda$ expansion terms to the action is at a higher order than 
$O(p^4)$ in the chiral counting which is not considered in the present work.}
\begin{eqnarray}
S=-\kappa\frac 12\int \mbox{tr} \left(K(z)^{-1/3}F_{\mu\nu}^2+ 2K(z)F_{\mu z}\right) d^4x \, dz .
\nonumber \\
\label{SSaction}
\end{eqnarray}
Here the KK mass $M_{KK}$ is set equal to 1 but will be recovered in actual calculations.
The warping factor is reduced in a series of approximations to the simple form
\begin{equation}
K(z)= 1+z^2.
\end{equation}
Setting $K(z)=1$, one arrives at the flat space.
This will be considered below in connection with the BPS Skyrmion.
The topological CS term, being background-independent, is the same as given in Eq.~(\ref{CS}).

The structure of baryons as instantons in the 5D YM action (\ref{SSaction}) plus the CS term was worked out
in Refs.~\cite{HRYY07-HRYY07a-HRYY07b,HSS08}.
They correspond to the Skyrmions in the presence of the pion and the infinite tower of vector mesons.
What we would like to do is to compare the truncated models where certain vector mesons are omitted to this
infinite-tower structure.
One can then learn how the vector mesons contribute in the presence of the warping.
To do this we integrate out all vector mesons in the tower except for the lowest, $\rho$ and $\omega$.
We shall call the resulting Lagrangian HLS$_1$.

How to integrate out the tower preserving hidden local symmetry of the vector mesons that are being
eliminated was worked out in Ref.~\cite{HMY10} and the full expression valid to the chiral order $O(p^4)$
needed for the exact Skyrmion calculation to that order is listed in Ref.~\cite{MOYH12}.
In a nutshell, the idea is as follows.
When the YM action is KK-decomposed by dimensional reduction to an infinite tower of both vector and
axial-vector mesons, one can rewrite the resulting action in terms of a tower of hidden local symmetric fields.
One then integrates out $n$ HLS fields with $n>1$ preserving hidden local symmetry for the remaining $n=1$
fields that are to be treated as the relevant degrees of freedom.
As shown in Ref.~\cite{HMY10}, this turns out to be equivalent to setting the mass eigenstate fields
-- but not hidden local fields -- for $n>1$  to zero. 
It should be noticed that the ``integrating out" adopted here is different from the ``naive truncation" which 
violates the chiral invariance, as explained in detail in Ref.~\cite{HMY10}. 
Actually, in the procedure, the equations of motion for the higher modes are solved based on the order
 counting of the derivative expansion, and the solutions are substituted back into the action. 
To the same chiral order $O(p^4)$, there are of course one-loop graphs that give non-local contributions 
but they are suppressed by $N_c$.
The power of this integrating-out procedure is that hidden local symmetry allows to do a systematic power counting
in the sense of chiral perturbation theory.
This is not just a ``philosophical advantage" but has a predictive power when applied to vector mesons in medium
where the masses can go to zero in the chiral limit~\cite{BR91,HY03a}.

To $O (p^4)$ in the large $N_c$ limit, when the external sources are switched off,
the HLS$_1$ Lagrangian~\cite{HY03a} is
\begin{equation}
\mathcal{L}_{\rm HLS_1} = \mathcal{L}_{(2)} + \mathcal{L}_{(4)y} +
\mathcal{L}_{(4)z} + \mathcal{ L}_{\rm an},
\label{eq:lagrhlstsruct}
\end{equation}
where the subscript $(n)$ represents the power $O(p^n)$ and
\begin{eqnarray}
\mathcal{L}_{\rm (2)} & = & f_\pi^2 \, \mbox{tr} [\hat{\alpha}_{\perp\mu}^{} \hat{\alpha}_{\perp}^{\mu}]
+ a f_\pi^2 \, \mbox{tr} [\hat{\alpha}_{\parallel\mu}^{} \hat{\alpha}_{\parallel}^{\mu}]
- \frac{1}{2g^2} \mbox{tr} [V_{\mu\nu} V^{\mu\nu}],
\nonumber \\
\label{eq:lagrp2}\\
%%%%%%%%%%%%%%%%%%%%%%%%%%%%%%%%%%%%%%%%%%%%%%%%%%%%%%%%%%%%%%%%%%%%%%%%%%%%%
\mathcal{ L}_{(4)y} &=& \sum_{i=1}^{9} y_i \, \mathcal{L}^4_i,
\label{eq:lagrp4y} \\
\mathcal{L}_{(4)z} & = & i z_4^{}  \, \mbox{tr} \left[ V_{\mu\nu}
\hat{\alpha}_\perp^\mu \hat{\alpha}_\perp^\nu \right]
+ i z_5^{}\,  \mbox{tr} [ V_{\mu\nu} \hat{\alpha}_\parallel^\mu
\hat{\alpha}_\parallel^\nu ]  ,
\label{eq:lagrp4z}\\
%%%%%%%%%%%%%%%%%%%%%%%%%%%%%%%%%%%%%%%%%%%%%%%%%%%%%%%%%%%%%%%%%%%%%%%%%%
\mathcal{L}_{\rm an} & = & \frac{N_c}{16\pi^2} \int_{M^4}
\sum_{i=1}^3 c_i^{} \, \mathcal{L}_i ,
\label{Lag:Anom}
\end{eqnarray}
where
\begin{eqnarray}
%%%%%%%%%%%%%%%%%%%%%%%%%%%%%%%%%%%%%%%%%%%%%%%%%%%%%%%%%%
\mathcal{L}_1 & = & i \, \mbox{tr} \left[\hat{\alpha}_{L}^3 \hat{\alpha}_{R}^{}
- \hat{\alpha}_{R}^3 \hat{\alpha}_{L}^{} \right] ,
\\
\mathcal{ L}_2 & = & i \, {\rm tr} \left[\hat{\alpha}_{L}^{} \hat{\alpha}_{R}^{}
\hat{\alpha}_{L}^{} \hat{\alpha}_{R}^{} \right] , \\
\mathcal{L}_3 & = & {\rm tr}\left[ F_{V} \left( \hat{\alpha}_{L}^{}
\hat{\alpha}_{R}^{} - \hat{\alpha}_{R^{}} \hat{\alpha}_{L}^{} \right) \right] .
\end{eqnarray}
Here, $f_\pi$ is the pion decay constant.
The axial-vector field $\hat{\alpha}_{\perp\mu}$ and vector field $\hat{\alpha}_{\parallel\mu}$ are defined as
\begin{eqnarray}
\hat{\alpha}_{\perp\mu}^{} &=& \frac{1}{2i} \left( D_\mu \xi_R^{} \xi_R^\dagger
- D_\mu \xi_L^{} \xi_L^\dagger \right),
\nonumber \\
\hat{\alpha}_{\parallel\mu}^{} &=& \frac{1}{2i} \left( D_\mu \xi_R^{} \xi_R^\dagger
+ D_\mu \xi_L^{} \xi_L^\dagger \right),
\end{eqnarray}
where
\begin{equation}
D_\mu \xi_{L,R}^{} = \left( \partial_\mu - ig V_\mu \right) \xi_{L,R}^{}
\end{equation}
with the vector meson field $V_\mu$.
The field strength tensor of the vector meson field is $V_{\mu\nu}$
and $F_V$ is its 1-form notation, $F_V = dV - i V^2$.
We also define
\begin{equation}
\hat{\alpha}_L^{} = \hat{\alpha}_\parallel^{} - \hat{\alpha}_{\perp}^{}, \qquad
\hat{\alpha}_R^{} = \hat{\alpha}_\parallel^{} + \hat{\alpha}_{\perp}^{}.
\end{equation}
The $\mathcal{L}^4_i $'s in Eq.~(\ref{eq:lagrp4y}) are independent $O(p^4)$ (hidden) gauge invariant terms
built with the covariants $\hat{\alpha}_\perp^\mu $ and $ \hat{\alpha}_\parallel^\mu$, and
their explicit expressions can be found in Ref.~\cite{HY03a}.
What we have here is the most general expression of the HLS$_1$ Lagrangian to $O (p^4)$
relevant to the problem at issue.
It contains 17 parameters. In standard chiral perturbation theory, these constants will have to be fixed
from experimental or theoretical information in the meson sector.
This is, however, not feasible at present because of the lack of enough information.
What makes the calculation performed in Ref.~\cite{MOYH12} feasible is that \textit{all} the parameters
are given in terms of the two parameters $f_\pi$ and $\lambda$ that are determined in the meson sector
by the pion decay constant and the mass of the $\rho$ meson in the hQCD model.
It is this feat that we shall exploit in what follows.

If we integrate out the entire tower of vector mesons, namely, the lowest vector mesons as well
in Eq.~(\ref{eq:lagrhlstsruct}), then we wind up with the Skyrme model with pions only,
\begin{eqnarray}
\mathcal{L}_{\rm ChPT} & = &
f_\pi^2 \mbox{tr}\, [\alpha_{\perp\mu}\alpha_{\perp}^{\mu}]
+\frac{1}{2 e^2} \mbox{tr}\left([\alpha_{\perp \mu},\alpha_{\perp \nu}]
[\alpha_{\perp}^{ \mu},\alpha_{\perp}^{\nu}] \right)
\nonumber\\
& = & \frac{f_\pi^2}{4} \mbox{tr} \left( \partial_\mu U\partial^\mu U^\dagger \right)
+\frac{1}{32e^2} \mbox{tr} [U^\dagger\partial_\mu U,U^\dagger\partial_\nu U]^2
\nonumber\\
\end{eqnarray}
with
\begin{equation}
\frac{1}{2 e^2} =  \frac{1}{2g^2} - \frac{z_4^{}}{2} - \frac{y_1^{} - y_2^{}}{4}.
\label{skyrmeterm}
\end{equation}
We should note that there are no other quartic-order terms than the Skyrme term.
A term of the form $\frac{y_1^{} + y_2^{}}{4}\, \mbox{tr} [\{\alpha_{\perp \mu},\alpha_{\perp \nu}\}
\{\alpha_{\perp}^{ \mu},\alpha_{\perp}^{\nu}\}]$, where the curly bracket represent the anti-commutator,
is present but it vanishes because the coefficient is exactly zero by cancellation in the SS model.
This is not the case in general.
However, it is noteworthy that in chiral perturbation theory for $\pi$-$\pi$ scattering,
this term, while nonzero, is small compared with the Skyrme term~\cite{BCG94-DP00}.
Note also that integrating out the vectors from HLS$_1$ term brings in corrections to
what one would obtain when all the vector fields are set equal to zero.
The second and third terms of Eq.~(\ref{skyrmeterm}) result from terms involving vector mesons
when the latter are integrated out.
It turns out that $\left( \frac{z_4^{}}{2} + \frac{y_1^{} - y_2^{}}{4} \right) >0$, so the constant $1/e$ is less than
$1/g$ that one gets by sending the mass of the $\rho$ meson to infinity.

\section{\boldmath Results of HLS$_1$ Skyrmion in a warped space}

\subsection{Instanton}

The ``reference result" to which comparison is to be made is that of the instanton description
with the SS model obtained in Refs.~\cite{HRYY07-HRYY07a-HRYY07b,HSSY07,HSS08}.
For the parameters $f_\pi=92.4$~MeV and $\lambda=17$ fixed in the meson sector~\cite{SS04a-SS05},
the mass of the instanton is~\cite{HRYY07-HRYY07a-HRYY07b}%
\footnote{We give  approximate numerical values with the understanding that the parameters fixed
in the meson sector that we use are highly approximative. Precise values for the HLS$_1$\ Skyrmions are found in Ref.~\cite{MOYH12}.}
\begin{equation}
M_{\rm instanton}\simeq 1800\  \mbox{MeV}.
\label{infinite1}
\end{equation}
This corresponds to the mass of a Skyrmion in the infinite tower of vector mesons in a warped space
and the Chern-Simons term.
The collective quantization gives the $\Delta$-$N$ mass difference that arises at $O(1/N_c)$ as~\cite{HSSY07}
\begin{equation}
\Delta M \equiv m_\Delta^{} -m_N^{} \approx 570\ \mbox{MeV},
\label{infinite2}
\end{equation}
where $m_{\Delta,N}^{}$ is the mass that contains the rotational $1/N_c$ contribution.

In the above estimates, the KK mass $M_{KK}$ which sets the scale or cutoff was taken to be
$M_{KK} \simeq 950$~MeV as fixed by the two parameters in the meson sector~\cite{SS04a-SS05}.
Both the mass in Eq.~(\ref{infinite1}) and the splitting in Eq.~(\ref{infinite2}) are much too big compared
with the experimental data.
As noted in Ref.~\cite{HSSY07}, were we to reduce $M_{KK}$ to $\sim 500$~MeV, we would get
$\sim 950$~MeV for the soliton mass and $\sim 300$~MeV for the $\Delta$-$N$ mass difference,
both consistent with experiments.
This is similar to the reduced \textit{effective} $f_\pi$ first used in Ref.~\cite{ANW83} for the Skyrme model.
How to reconcile results with Nature by implementing a dilaton scalar degree of freedom will be discussed in the last section.

\subsection{\boldmath HLS$_1$ Skyrmion with $\rho$, $\omega$ and $\pi$}

Next we consider integrating out \textit{all} vector mesons except for
the lowest vector mesons $\rho$ and $\omega$.
The resulting Lagrangian is given in Eq.~(\ref{eq:lagrhlstsruct}).
What distinguishes this Lagrangian from the conventional -- and truncated -- HLS Lagrangian
used in the past is that it is complete in chiral order to $O (p^4)$ in both the normal
and anomalous components of the Lagrangian and furthermore there are no unknown parameters.
In the past, the anomalous part of the Lagrangian -- referred to as ``homogenous Wess-Zumino (hWZ)" term --
was often approximated by one term proportional to $\omega_\mu B^\mu$ where $B_\mu$ is the baryon number current.
This form requires assuming $m_\rho\rightarrow \infty$ in the hWZ term which is not consistent with the notion
that the $\rho$ mass is of the same chiral order as the pion mass indispensable for hidden local symmetric approaches.

There is one more important aspect of the HLS$_1$ soliton we are considering that needs to be signaled
and that is that the properties of the soliton of this HLS$_1$ model should have no $a$ dependence that appears
in Eq.~(\ref{eq:lagrp2}).
In the holographic setting, $a$ is linked to the normalization of the lowest eigenvalue $\lambda_1$ for $\psi_1(z)$
and physical quantities of the baryon should be independent of $a$.
The proof for this observation is given in detail in Ref.~\cite{MOYH12}.
In the standard -- or boundary or gauge sector -- HLS theory~\cite{BKY88,HY03a}, $a$ is defined in the range
$1\leq a\leq 2$.
It takes $a \simeq 2$ in free space and $a \simeq 1$ in hadronic medium, i.e., at high temperature and/or density~\cite{HY03a}.

We now quote the results of the involved calculation of Ref.~\cite{MOYH12} for the soliton mass and collective quantization,
and comment on their implications.
Denoting by $\mathcal{M}$ those degrees of freedom left un-integrated out, we have
\begin{itemize}
\item $\mathcal{M} =\pi,\rho,\omega$:
\begin{eqnarray}
&& M_{\rm HLS_1} (\pi,\rho,\omega) \approx 1184 \mbox{ MeV}, \nonumber\\
&& \Delta M \equiv m_\Delta^{} -m_N^{} \approx 448 \mbox{ MeV}.
\end{eqnarray}
Note that the soliton mass is of $O (N_c)$ while  $\Delta M$ is of $O (1/N_c)$.
\item $\mathcal{M} =\pi,\rho$: Now we integrate out the $\omega$ meson and find
\begin{eqnarray}
&& M_{\rm HLS_1} (\pi,\rho) \approx 835 \mbox{ MeV}, \nonumber\\
&& \Delta M \approx 1707 \mbox{ MeV}.\label{44}
\end{eqnarray}
\item $\mathcal{M} =\pi$: Finally integrating out the last vector meson $\rho$ winding up with the Skyrme model, one gets
\begin{eqnarray}
&& M_{\rm HLS_1} (\pi) \approx 922 \mbox{ MeV},\nonumber\\
&& \Delta M \approx 1014 \mbox{ MeV}.\label{45}
\end{eqnarray}
\end{itemize}

What transpires here can be summarized as follows:
As the isovector vector mesons are added, the soliton mass decreases
as in the BPS case while the $\Delta M$ increases.
On the other hand, when the isoscalar vector meson is added,
the soliton mass \textit{increases} while $\Delta M$ \textit{decreases}.
One can easily understand this inverse correlation between the soliton mass
and the $\Delta$-$N$ mass splitting by looking
at what happens when \textit{all} vector mesons are integrated out giving the Skyrme model.
Because of the reduction of $1/e^2$ by the second and third terms in Eq.~(\ref{skyrmeterm}), the soliton mass gets reduced.
But it increases the $\Delta$-$N$ splitting which goes proportional to $e$.
This problem is avoided in Ref.~\cite{ANW83} by reducing \textit{both} $f_\pi$ and $e$.
We suggest that this is intricately correlated with the axial-vector coupling constant $g_A^{}$.
Keeping $f_\pi$ at its physical value and adjusting $e$ to give $g_A^{} = 1.26$ would lead to the soliton mass
$M_{\rm Skyrme} \sim 1500$~MeV.%
\footnote{See Ref.~\cite{NRZ} for a discussion on this matter.}

Two points are worth noticing here. One is that while there is a tendency of flow to conformality
in the soliton mass with the isovector vector mesons even with a warped space, the isoscalar vector mesons
\textit{strongly} counter this tendency.
On the other hand, the $\omega$ meson that plays a crucial role in the repulsion in nucleon interactions
reduces an unrealistically large $\Delta$-$N$ splitting from that without the $\omega$ meson.
This feature is generic independent of the background warping as we shall see below with BPS Skyrmions.
The striking influence of the $\omega$ meson in the soliton structure was also observed in dense medium
described by HLS Lagrangian treated in terms of crystals in Ref.~\cite{PRV03}.
The connection between these diverse phenomena, i.e., the universal hard-core repulsion, the apparent
obstruction to conformal flow and the $\Delta$-$N$ splitting etc. is a deep open problem in nuclear physics.

 We now suggest that what is happening here with $g_A^{}$ can be exploited to remove the defects in
the instanton results (\ref{infinite1}) and (\ref{infinite2}), both of which are too big.
As noted in Refs.~\cite{HRYY07-HRYY07a-HRYY07b}, when an $O (N_c^0)$ correction is suitably made
to the axial coupling constant in the Sakai-Sugimoto model, one gets $g_A^{} = \frac{g_A^0}{3} N_c (1+2/N_c)$
where $g_A^0$  comes out to be $\sim 0.75$, so for $N_c=3$,  one gets $g_A \approx 1.25$
consistent with the experimental value $1.27$.

 Up to date, there has  been no derivation of this $O(N_c^0)$ Casimir contribution in holographic models.
It is tantamount to making $1/N_c$ corrections and this task remains unresolved in holographic approaches,
so is ignored in the string theory community.
However this $O (1)$ term comes out naturally in the large $N_c$ counting in the non-relativistic quark model
as well as in the Skyrmion quantization.
In a similar vein, we note that the instanton mass is of $O(N_c)$ whereas the splitting $\Delta M$ is of
$O(1/N_c)$.
The $O(1)$ Casimir energy is glaringly missing.
Just as the $O(1)$ term is important for $g_A^{}$, such an $O(1)$ term could also be important
for the baryon mass.
The Casimir calculation is notoriously difficult to perform given that we have a non-renormalizable theory
but there is nothing that suggests that it should not be there.
In fact, the presently available estimate in the Skyrme model, though admittedly very rough, does indeed
give an attractive Casimir contribution of order $\sim - 500$~MeV, going in the right direction with a correct
order of magnitude~\cite{NRZ}.
As we will discuss below (in the last section), this defect could be remedied by implementing scalar degrees
of freedom missing in the holographic model. Such scalars could contribute the missing $O(1)$ effects.

\section{BPS Skyrmion and the Chern-Simons term}

We learned from the work of Sutcliffe~\cite{sutcliffe11,sutcliffe10} that the Skyrmion in the flat space
5D YM action, i.e., BPS Skyrmion, has the potentially important feature that the more vector mesons
in the infinite tower in 4D are implemented, the closer the Skyrmion mass approaches the BPS mass $8\pi^2B$, that is, the theory flows to conformal theory.
In this consideration the Chern-Simons term which encapsulates chiral anomaly has not been
taken into account.
The CS term is background-independent and hence should be independent of the warping.
We show that the CS term plays a qualitatively similar role in the BPS Skyrmion model
as in the HLS$_1$ model with the warped background.

Using our energy unit, we have the BPS mass $M_{\rm BPS} \approx \frac{\lambda N_c}{27\pi} M_{KK}
\approx 559$~MeV~\cite{HRYY07-HRYY07a-HRYY07b}%
\footnote{Here we used $\lambda=16.66, M_{KK} =948$~MeV determined from our inputs.}
in agreement with Ref.~\cite{sutcliffe10}.
When the CS term contribution is added, we get $M_{\rm BPS-CS} = M_{\rm BPS} + \sqrt{\frac{2}{15}} N_c M_{KK}
\approx 1038$~MeV.
In looking at the cases where the tower is integrated out, we will follow the same procedure
as in the case of the SS model. We will first integrate out all except the lowest vector mesons
$\rho$ and $\omega$ and the pion, then integrate out  the $\omega$ and then finally the $\rho$.
For the given $\mathcal{M}$, the results are:
\begin{itemize}
\item $\mathcal{M}=\pi, \rho,\omega$:
\begin{eqnarray}
M_{\rm BPS}(\pi,\rho,\omega)&\approx & 1162\ {\rm MeV}, \nonumber \\
\Delta M_{\rm BPS} (\pi,\rho,\omega) &\approx& 456\ {\rm MeV}.
\end{eqnarray}
\item $\mathcal{M}=\pi, \rho$:
\begin{eqnarray}
M_{\rm BPS}(\pi,\rho)&\approx & 577\ {\rm MeV}, \nonumber \\
\Delta M_{\rm BPS} (\pi,\rho) &\approx& 4541\ {\rm MeV}.\label{47}
\end{eqnarray}
\item $\mathcal{M}=\pi$:
\begin{eqnarray}
M_{\rm BPS}(\pi)&\approx & 673\ {\rm MeV}, \nonumber \\
\Delta M_{\rm BPS} (\pi) &\approx& 2611\ {\rm MeV}.\label{48}
\end{eqnarray}
\end{itemize}
Although in magnitude they are different, one observes qualitatively the same tendency
in the opposing effect in the soliton mass and the mass splitting as in the HLS$_1$ model:
The $\omega$ meson blocks the flow to the conformal fixed point while reducing the $\Delta$-$N$ mass splitting.

%\section{Application to dense matter}
\section{Discussions}

In this Section we briefly summarize our findings in the single-baryon sector and 
then make a few comments on their implications on dense matter relevant for the 
physics of compact stars, the main objective of the series of work in progress.

In the large $N_c$ and large $\lambda$ limit, the Skyrmion embedded in the tower 
of isovector vector mesons as described by a 5D YM action without the CS term 
(which is absent at the leading order in $\lambda$) flows to a BPS instanton as more 
vector mesons are included.
The interaction gets weaker and the size becomes smaller.
This tendency however gets blocked at the next order in $\lambda$, namely at $O(\lambda^0)$,
by the presence of the $\omega$ meson present in the CS term.
The effect of the $\omega$ meson is two-fold.
It increases the soliton mass way above the empirical nucleon mass and decreases its size way 
below the empirical size~\cite{MOYH12}.
This correlation is not difficult to understand.
What is surprising however is what happens with the hyperfine splitting $\Delta M$ between
the ground state $N$ and its rotation excitation $\Delta$.
It comes out to be more than 5 times the empirical value in the absence of the $\omega$
(lodged in the CS term) and gets reduced by a factor of more than $\sim 3$ in its presence.
As mentioned, these drastic effects of the $\omega$ at the next-to-leading order in $1/\lambda$,
points to a possible importance of both $1/N_c$ and $1/\lambda$ corrections in the baryon structure.
It has been observed in the standard Skyrme model that some, if not all, of the problems can be resolved
by $1/N_c$ corrections -- via Casimir energy -- to the mass and to the axial coupling constant $g_A^{}$.
In terms of hidden local symmetry Lagrangian, there has been an attempt, with some success,
to remedy these difficulties by implementing a scalar degree of freedom, dilaton, associated with
the QCD trace anomaly~\cite{LR09}.
The dilaton provides an attraction that significantly compensates the $\omega$ repulsion, 
thereby reducing the mass.
The basic difficulty in the bulk-sector model, however, is that there is no way known to introduce
a low-mass scalar that would simulate the attraction required.%
\footnote{There is a scalar attraction in the Sakai-Sugimoto model but the scalar is much too heavy
to be identified with the scalar that is needed for the single baryon as well as in nuclear matter, 
discussed below~\cite{KS11}.}

One of the most striking -- and puzzling -- observations made in this paper is the role of 
the $\omega$ meson in the $\Delta$-$N$ mass splitting. 
It involves both the large $N_c$ and large $\lambda$ approximations. 
The effect in question appears both in the warped space, (\ref{44})-(\ref{45}),  and in the flat space, 
(\ref{47})-(\ref{48}). 
That the $O(1/N_c)$ terms associated with the mass splitting are an order of magnitude greater 
than the $O(N_c)$ terms of the soliton mass suggest either that the large $N_c$ expansion and/or 
large $\lambda$ expansion make no sense whatsoever or the role of $\omega$ meson is 
not at all understandable, or both. 
This observation appears to crack wide open the issue of the {\it right} degrees of freedom that 
should figure in effective Lagrangians for the solitonic approach to baryons.

The prominent effects of the $\omega$ meson in the baryon structure observed in this paper 
must be correlated also with the role it plays in nuclear interactions.
In the effective field theory framework modeling QCD, it is well established that the vector, 
$\omega$, degree of freedom is essential for the stability of nuclear matter.
In a mean-field theoretic description, it is the balance between the $\omega$ repulsion and the
scalar attraction of a range comparable to that of the $\omega$ that provides the nuclear saturation.
Thus very two effects that have not been handled in the bulk sector must play an important role in nuclear physics,
namely $1/\lambda$ and $1/N_c$ corrections and a low-mass scalar (of $\sim 600$~MeV).
What happens to the balance between the attraction and the repulsion when the system is squeezed
to high density as in compact stars is therefore totally unknown.

In a recent work based on renormalization group property of hidden local symmetric Lagrangian taken
at $O(p^2)$ in baryonic medium, it was found that as density approaches the chiral restoration density,
the vector-meson--nucleon coupling should go to zero at some density referred to
as ``dilaton limit fixed point"~\cite{PLRS11,SLPR11}.
This would mean that the $\omega NN$ coupling should decrease as density increases.
This turns out to bring havoc to nuclear matter at a density $n \gtrsim 2n_0$ as it would make
the neutron-star equation of state (EoS) much too soft to support the observed 2-solar mass star~\cite{DKLR12}.
Assuming that this consideration applies also to the bulk-sector theory,  a way out of this difficulty
might be the intervention of the tower of isoscalar vector mesons which become important
as the lowest $\omega$ gets suppressed.
This would make the nature of short-range repulsion basically different from the standard interpretation
in terms of $\omega$-exchange many-body forces.

An extension of the model so far studied  to dense matter is to put the Skyrmions considered in this paper
on crystal lattice and determine where in density a Skyrmion (or instanton) transforms to two
half-Skyrmions~\cite{LPR11,LR11} (or half-instantons/dyons~\cite{RSZ09}).
This is important in calculating the EoS for compact-star matter as shown in Ref.~\cite{DKLR12}.
In doing so, the missing ingredient is the scalar degree of freedom which figures importantly in the
previous studies with truncated HLS Lagrangian~\cite{PLRS11,SLPR11,LPRV03}.
Although putting a scalar of a mass relevant to nuclear matter at high density into the SS model is still unknown,
the indication from the Skyrmion case~\cite{LPRV03}  is that the density at which the change-over occurs
is highly insensitive to the mass of the scalar.
What is relevant then would be the vector mesons considered in this paper and what was found here
is expected to be of importance to the problem.

\subsection*{Acknowledgments}

The work reported here was partially supported by the WCU project of Korean
Ministry of Education, Science and Technology (R33-2008-000-10087-0).
The work of M.H. and Y.-L.M. was supported in part by Grant-in-Aid for
Scientific Research on Innovative Areas (No. 2104) ``Quest on New
Hadrons with Variety of Flavors'' from MEXT.
Y.-L.M. was supported in part by the
National Science Foundation of China (NNSFC) under Grant No.
10905060.
The work of M.H. was supported in part by the Grant-in-Aid for
Nagoya University Global COE Program
``Quest for Fundamental Principles in the Universe: from Particles
to the Solar System and the Cosmos'' from MEXT, the JSPS
Grant-in-Aid for Scientific Research (S) $\sharp$ 22224003, (c)
$\sharp$ 24540266.
The work of Y.O. and G.-S.Y. was supported in part by the Basic Science Research
Program through the National Research Foundation of Korea (NRF) funded by
the Ministry of Education, Science and Technology (Grant \mbox{No.} 2010-0009381).
A part of this work was discussed at the International Workshop on ``Dense Strange Nuclei and
Compressed Baryonic Matter" (Dense11) at YITP, Kyoto University.


\begin{thebibliography}{10}

\bibitem{BKY88}
M.~Bando, T.~Kugo, and K.~Yamawaki,
\newblock Phys. Rep. \textbf{164}, 217 (1988).
%%CITATION = PRPLC,164,217;%%

\bibitem{HY03a}
M.~Harada and K.~Yamawaki,
\newblock Phys. Rep. \textbf{381}, 1 (2003).
%%CITATION = HEP-PH 0302103;%%

\bibitem{SS03}
D.~T. Son and M.~A. Stephanov,
\newblock Phys. Rev. D \textbf{69}, 065020 (2004).
%%CITATION = HEP-PH 0304182;%%

\bibitem{SS04a-SS05}
T.~Sakai and S.~Sugimoto,
\newblock Prog. Theor. Phys. \textbf{113}, 843 (2005);
%%CITATION = HEP-TH 0412141;%%
%
%\bibitem{SS05}
%T.~Sakai and S.~Sugimoto,
\newblock Prog. Theor. Phys. \textbf{114}, 1083 (2005).
%%CITATION = HEP-TH 0507073;%%

\bibitem{LPMRV03}
H.-J. Lee \textit{et~al.\/},
\newblock Nucl. Phys. A \textbf{723}, 427 (2003).
%%CITATION = HEP-PH 0302019;%%

\bibitem{meissner} 
U.~G.~Meissner,
\newblock Phys. Rept.  \textbf{161}, 213 (1988).  
%%CITATION = PRPLC,161,213;%%
 
\bibitem{park-vento} 
B.-Y. Park and V. Vento, 
``Skyrmion approach to finite density and temperature," in 
\textit{The Multifaceted Skyrmions} (World Scientific, Singapore, 2010) 
edited by G.~E. Brown and M.~Rho.
%%CITATION = ARXIV:0906.3263;%%
     
\bibitem{manton-sutcliffe} 
R.~A. Battye, N.~S. Manton, and P.~M. Sutcliffe, 
``Skyrmions and nuclei," in \textit{The Multifaceted Skyrmions} 
(World Scientific, Singapore, 2010) edited by G.~E. Brown and M.~Rho.
 
\bibitem{sutcliffe11}
P.~Sutcliffe,
\newblock JHEP \textbf{1104}, 045 (2011).
%%CITATION = ARXIV:1101.2402;%%

\bibitem{HRYY07-HRYY07a-HRYY07b}
D.~K. Hong, M.~Rho, H.-U. Yee, and P.~Yi,
\newblock Phys. Rev. D \textbf{76}, 061901 (2007);
%%CITATION = HEP-TH 0701276;%%
%
%\bibitem{HRYY07a}
%D.~K. Hong, M.~Rho, H.-U. Yee, and P.~Yi,
\newblock JHEP \textbf{0709}, 063 (2007);
%%CITATION = ARXIV:0705.2632;%%
%
%\bibitem{HRYY07b}
%D.~K. Hong, M.~Rho, H.-U. Yee, and P.~Yi,
\newblock Phys. Rev. D \textbf{77}, 014030 (2008).
%%CITATION = ARXIV:0710.4615;%%

\bibitem{HSSY07}
H.~Hata, T.~Sakai, S.~Sugimoto, and S.~Yamato,
\newblock Prog. Theor. Phys. \textbf{117}, 1157 (2007).
%%CITATION = HEP-TH 0701280;%%

\bibitem{KMS12}
V.~Kaplunovsky, D.~Melnikov, and J.~Sonnenschein, arXiv:1201.1331.
%%CITATION = ARXIV:1201.1331;%%

\bibitem{sutcliffe10}
P.~Sutcliffe,
\newblock JHEP \textbf{1008}, 019 (2010).
%%CITATION = ARXIV:1003.0023;%%

\bibitem{AM89}
M.~F. Atiyah and N.~S. Manton,
\newblock Phys. Lett. B \textbf{222}, 438 (1989).
%%CITATION = PHLTA,B222,438;%%

\bibitem{Skyrme62}
T.~H.~R. Skyrme,
\newblock Nucl. Phys. \textbf{31}, 556 (1962).
%%CITATION = NUPHA,31,556;%%

\bibitem{JR83}
A.~D. Jackson and M.~Rho,
\newblock Phys. Rev. Lett. \textbf{51}, 751 (1983).
%%CITATION = PRLTA,51,751;%%

\bibitem{ANW83}
G.~S. Adkins, C.~R. Nappi, and E.~Witten,
\newblock Nucl. Phys. B \textbf{228}, 552 (1983).
%%CITATION = NUPHA,B228,552;%%

\bibitem{HSS08}
K.~Hashimoto, T.~Sakai, and S.~Sugimoto,
\newblock Prog. Theor. Phys. \textbf{120}, 1093 (2008).
%%CITATION = ARXIV:0806.3122;%%

\bibitem{AHI12}
For a recent review,
S.~Aoki, K.~Hashimoto, and N.~Iizuka, arXiv:1203.5386.
%%CITATION = ARXIV:1203.5386;%%

\bibitem{PRV03}
B.-Y. Park, M.~Rho, and V.~Vento,
\newblock Nucl. Phys. A \textbf{736}, 129 (2004).
%%CITATION = HEP-PH 0310087;%%

\bibitem{MOYH12}
Y.-L. Ma, G.-S. Yang, Y.~Oh, and M.~Harada, 
``Skyrmions with vector mesons in the hidden local symmetry approach," 
\newblock (to appear).

\bibitem{HMY10}
M.~Harada, S.~Matsuzaki, and K.~Yamawaki,
\newblock Phys. Rev. D \textbf{82}, 076010 (2010).
%%CITATION = ARXIV: 1007.4715;%%

\bibitem{BR91}
G.~E. Brown and M.~Rho,
\newblock Phys. Rev. Lett. \textbf{66}, 2720 (1991).
%%CITATION = PRLTA,66,2720;%%

\bibitem{BCG94-DP00}
J.~Bijnens, G.~Colangelo, and J.~Gasser,
\newblock Nucl. Phys. B \textbf{427}, 427 (1994);
%%CITATION = HEP-PH 9403390;%%
%
%\bibitem{DP00}
for a discussion on this point, see
D.~Diakonov and V.~\mbox{Yu}. Petrov,
\newblock Nucleons as chiral solitons,
\newblock in \textit{At the Frontier of Particle Physics, Vol 1.}, edited by
  M.~Shifman, pp. 359--415, World Scientific, Singapore, 2001.

\bibitem{NRZ}
M.~A. Nowak, M.~Rho, and I.~Zahed,
\newblock \textit{Chiral Nuclear Dynamics} (World Scientific, Singapore, 1996).

\bibitem{LR09}
H.~K. Lee and M.~Rho,
\newblock Nucl. Phys. A \textbf{829}, 76 (2009).
%%CITATION = ARXIV:0902.3361;%%

\bibitem{KS11}
V.~Kaplunovsky and J.~Sonnenschein,
\newblock JHEP \textbf{1105}, 058 (2011).
%%CITATION = ARXIV:1003.2621;%%

\bibitem{PLRS11}
W.-G. Paeng, H.~K. Lee, M.~Rho, and C.~Sasaki,
\newblock Phys. Rev. D \textbf{85}, 054022 (2012).
%%CITATION = ARXIV:1109.5431;%%

\bibitem{SLPR11}
C.~Sasaki, H.~K. Lee, W.-G. Paeng, and M.~Rho,
\newblock Phys. Rev. D \textbf{84}, 034011 (2011).
%%CITATION = ARXIV:1103.0184;%%

\bibitem{DKLR12}
H.~Dong, T.~T.~S. Kuo, H.~K. Lee, and M.~Rho,  
``Half-Skyrmions and the Equation of State for Compact-Star Matter,''  
arXiv:1207.0429 [nucl-th].
%%CITATION = ARXIV:1207.0429;%%

\bibitem{LPR11}
H.~K. Lee, B.-Y. Park, and M.~Rho,
\newblock Phys. Rev. C \textbf{83}, 025206 (2011);
\newblock \textbf{84}, 059902(E) (2011).
%%CITATION = ARXIV:1005.0255;%%

\bibitem{LR11}
H.~K. Lee and M.~Rho,  ``Half-Skyrmion Hadronic Matter at High Density," arXiv:0905.0235.
%%CITATION = ARXIV:0905.0235;%%

\bibitem{RSZ09}
M.~Rho, S.-J. Sin, and I.~Zahed,
\newblock Phys. Lett. B \textbf{689}, 23 (2010).
%%CITATION = ARXIV:0910.3774;%%

\bibitem{LPRV03}
H.-J. Lee, B.-Y. Park, M.~Rho, and V.~Vento,
\newblock Nucl. Phys. A \textbf{741}, 161 (2004).
%%CITATION = HEP-PH 0307111;%%

\end{thebibliography}
\end{document}